# Neural Network Models for Stock Selection Based on Fundamental Analysis

Yuxuan Huang
Dept. of Electrical and Computer Engineering
Western University
London, Ontario, Canada
yhuan656@uwo.ca

Luiz Fernando Capretz
Dept. of Electrical and Computer Engineering
Western University
London, Ontario, Canada
lcapretz@uwo.ca

Danny Ho
NFA Estimation Inc.
Richmond Hill, Ontario, Canada
danny@nfa-estimation.com

*Abstract*—Application of neural network architectures for financial prediction has been actively studied in recent years. This paper presents a comparative study that investigates and compares feed-forward neural network (FNN) and adaptive neural fuzzy inference system (ANFIS) on stock prediction using fundamental financial ratios. The study is designed to evaluate the performance of each architecture based on the relative return of the selected portfolios with respect to the benchmark stock index. The results show that both architectures possess the ability to separate winners and losers from a sample universe of stocks, and the selected portfolios outperform the benchmark. Our study argues that FNN shows superior performance over ANFIS.

## I. INTRODUCTION

Stock trading is a process of buying and selling shares of publicly listed companies on a stock exchange platform, with millions of investors and traders from all over the world actively involved at any given time when the market is open. Stock market prediction is an extremely complex and difficult problem because there are simply too many factors and noises affecting the movement of the price. Many existing studies associated with stock market prediction support the well-known efficient market hypothesis (EMH), according to which the price of a stock at any given time reflects all information available about it and is therefore impossible to predict [1]. This problem remains a topic of interest among economists and researchers to this day [2]. There are three forms of EMH, based on the degree of stock market efficiency:

- **Weak form** EMH implies that the market efficiently reflects all past market information. The hypothesis assumes that past rates of return have no effect on future rates.
- **Semi-strong form** EMH implies that the market efficiently reflects all publicly available information. This hypothesis assumes that the stock price adjusts quickly to absorb new information. Semi-strong form incorporates weak form EMH.
- **Strong form** EMH implies that the market efficiently reflects all information, both public and private. This hypothesis assumes that no investor would be able to achieve above average returns even if he/she was given new information that is not available publicly. Strong form EMH incorporates weak form and semi-strong form EMH.

Recent studies which have explored using machine learning and soft computing techniques for stock prediction, have achieved results that challenge the weak and semi-strong form EMH [3]-[11]. However, most of these studies use historical price, technical indicators or investor sentiments as independent variables for model training and prediction. The main motivation of this research is to develop feed-forward neural network (FNN) and adaptive neural fuzzy inference system (ANFIS) models to resemble the decision-making process of investment experts based on a stock's fundamental financial ratios. Moreover, instead of simply predicting future absolute values of stocks, portfolios which consist of predicted winners or losers are selected and assessed. The portfolio selection mechanism resembles a more realistic approach to stock investment.

## II. RELATED WORK

The approaches investors and financial analysts use to predict stock market prices can be broadly classified into two types: fundamental analysis and technical analysis [12].

Fundamental analysis is based on a company's financial profile as well as macroeconomic data. Technical analysis on the other hand focuses solely on historical price and volume. Hu *et al.* [13] did a comprehensive literature review in 2015 and concluded that most existing studies which apply soft computing and neural network for financial prediction are based on technical analysis. An earlier literature review by Atsalakis *et al.* [14] shows the same pattern. Guresen *et al.* [3] explored using multi-layer perceptron (MLP) to predict the NASDAQ stock index. They treated the historical index price as a time series prediction problem and used Mean Square Error (MSE) and Mean Absolute Deviation (MAD) as evaluation criteria. Atsalakis *et al.* [4] applied ANFIS to predict the stock daily trend. The ANFIS model was compared with the FNN model and ANFIS achieved better performance in terms of hit ratio. Comparative studies which assessed different NN based architectures for stock prediction based on historical price movement were then conducted [5]-[7].

A few studies explored NN assisted stock selection based on fundamental analysis [8]-[11]. Shen and Tzeng[8] proposed a combined soft computing model for value stock selection. They concluded that their model could distinguish value stocks with satisfactory financial returns. Eakins and Stansell [9] applied FNN model for stock selection based on a set of fundamental financial ratios. They backtested their model for a 20-year period and achieved an investment return superior to that of benchmark stock index. Quah and Srinivasan [10] did a similar experiment using FNN for stock selection based on fundamental financial ratios. They also achieved above benchmark returns over their test period. In 2008, Quah [11] again compared FNN and ANFIS for stock selection based on fundamental analysis. Quah classified stocks into two classes based on their yearly return and converted the prediction problem into a classification problem. Eight years of quarterly data were used for training and 2 years for testing in Quah's study. His results show that ANFIS and FNN achieved comparable prediction results.

### III. NEURAL NETWORK ARCHITECTURES

#### A. Feed-forward Neural Network

FNN, or multi-layer perceptrons (MLP), is the simplest and most widely used form of neural network architecture. An FNN consists of at least three layers: an input layer, a hidden layer and an output layer. Each node is a neuron that uses a nonlinear activation function, except for the input neurons. The supervised learning technique of gradient descent is used for training. In the training process, the change of weight between two nodes is calculated as [15]:

$$\Delta w_{ji}(n) = -\gamma \frac{\delta \varepsilon(n)}{\delta v_j(n)} y_i(n) \quad (1)$$

where $\gamma$ is the learning rate, $\varepsilon$ is the error in the final output, $v_j$ is the induced local variable and $y_i$ is the output of the previous neuron.

#### B. Adaptive Neural Fuzzy Inference System

ANFIS is an instance of the more generic form of the Takagi-Sugeno-Kang (TSK) model. It replaces the fuzzy sets in the implication with a first order polynomial equation of the input variables [15]. The ANFIS system consists of rules in IF-THEN form. In general, there are five different layers in an ANFIS system. Layer 1 converts each input value to the outputs of its membership functions. Layer 2 calculates the firing strength of a rule. Layer 3 normalizes the firing strengths. Layer 4 consists of adaptive nodes with function defined as [16]:

$$O_i^4 = \overline{w_i}(p_i x + q_i y + r_i) \quad (2)$$

where $\overline{w_i}$ is the normalized firing strength from the previous layer and $(p_i x + q_i y + r_i)$ is a parameter in the node. Layer 5 sums all incoming signals and delivers a final output. The structure of a typical ANFIS is shown in Fig.1.

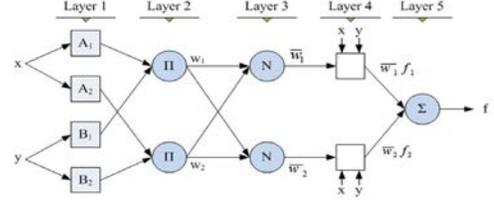

Fig.1. Structure of a ANFIS[15]

### IV. METHODOLOGY

For each stock, two models are developed for each algorithm. The experiments have been designed to evaluate the performance of the two algorithms in terms of the average quarterly portfolio return and the compounding portfolio return relative to the benchmark for the backtesting period.

#### A. Data Preprocessing

Sample stocks used for this experiment are chosen from the S&P 100 Index components. The index includes 102 leading U.S. stocks which represent about 51% of the market capitalization of the entire U.S. equity market [17]. There are two major reasons behind choosing the S&P 100 components as sample stocks. First, financial fundamental ratios for the S&P 100 stocks are relatively complete and large in terms of data volume. This is because these stocks are large-cap, and most of them were publicly listed relatively early in history. Second, the S&P 100 components are well balanced across different sectors, and we decide that the number of its components was suitable for the size of our sample stock universe. Because the composition of the S&P 100 index is frequently revisited, we decide to use its components as of December 2018 [17].

Historical financial data for each of the S&P 100 components were retrieved online in csv format [18]. These data are extracted from companies' SEC 10_Q filings published quarterly. In order to prepare the raw data for model fitting, we go through the following steps for data preprocessing:

*1) Feature Dropping*
Features with a high density of missing values consistently across stocks were also dropped.

*2) Trend Stationarization*
Our target variable in this research is quarterly relative returns, while many features from the raw dataset possess a clear global trend with respect to time. These features with global trends could hinder our supervised learning models' ability to generalize and provide reliable predictions. We therefore take the percentage change between consecutive observations for all features, which is calculated as follows:

$$\Delta y_t = \frac{(y_t - y_{t-1})}{y_{t-1}} \times 100\% \quad (3)$$

An example of trend stationarizing a feature is shown in Fig. 2.

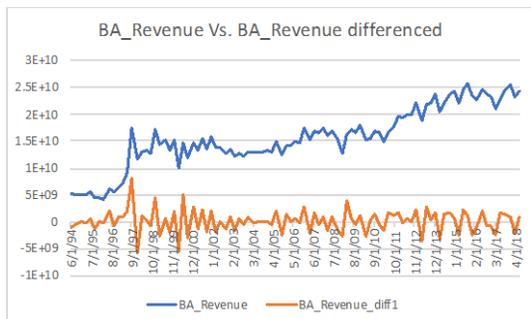

Fig. 2. Historical Quarterly Revenue for BA: the original data vs. after applying trend stationarization

### 3) Filling Missing Entries

Although features with many missing entries are dropped in step 1, there were still some sparsely located or missing entries. Mean substitution is used to replace these missing values with the average of their neighboring entries. Mean substitution is based on the assumption that the average of the neighboring observations in a time series is a good guess for a randomly selected observation [19].

### 4) Standardization

As the scales of features vary dramatically, standardization is applied to all features in order to improve the performance of our prediction models.

### 5) Fixing the Time Frame

We choose to use data from Q1 1996 to Q4 2017. This period provides 88 observations for each stock, as the financial data are published on a quarterly basis. Stocks with the earliest available observation later than Q1 1996 were dropped from the stock universe.

After data preprocessing, we end up with 21 features and 70 stocks. The 21 features are financial ratios such as $\Delta PE$, $\Delta Assets$, $\Delta liabilities$, etc. The only non-financial feature is the past quarter's relative return on price. Each stock has 88 observations, ranging from Q1 1996 to Q4 2017, with an interval of one quarter between two consecutive observations. The data are then partitioned into Train-Validation-Test sets in proportion of 60%-20%-20%. After model validation, the validation set is merged into the training set for final model training before generating testing results.

### B. Experiment

#### 1) Relative Return

In the experiment, a stock's quarterly relative return with respect to the Dow Jones Industrial Average (DJIA) is used as the target variable instead of simple absolute return. The DJIA is one of the most widely used U.S. stock market benchmarks. The relative return of a stock is the difference between its absolute return and the return of some benchmark. By subtracting overall market performance from the performance of each individual stock, we are able to filter out the factors affecting the broader market. In theory, using such a technique helps to reduce the complexity of the prediction problem and improve prediction performance and stability of our models.

#### 2) Portfolio Construction

After the simulated results are generated by the models for each of the 70 stocks, the stocks are then ranked by their predicted relative returns for the next quarter. Stocks with the top 30 predicted relative returns are chosen to construct a 'Buy' portfolio, and the bottom 30 stocks are put into a 'Sell' portfolio. In this experiment, we use equal-weight strategy for constructing portfolios. This means the hypothetical investment would be distributed equally across stocks in a portfolio. The portfolios are constructed for every quarter in the validation set during the model validation stage and the test set during the final testing stage.

#### 3) Model Validation

FNN and ANFIS models are both validated on the validation set. After validation, we settle on an FNN with a single hidden layer of 21 perceptrons. Rectifier and sigmoid were the activation functions for hidden layer and output layer respectively. The FNN uses mean squared error for loss function and Adam algorithm for optimization. For the ANFIS model, subtractive clustering with cluster influence range of 0.5 is used for defining membership functions and fuzzy rules.

### C. Evaluation

Both FNN and ANFIS models are tested using the test set which has 18 observations from Q3 2013 to Q4 2017 for each of the 70 stock. "Buy" and "Sell" portfolios are constructed for each of the 18 quarters in the testing period. Models are evaluated based on the mean and standard deviation of the quarterly relative return of the portfolio. Moreover, simulated compound relative return through the entire testing period is also computed for each model. The compound relative return is calculated based on the assumption that the investor redistributes his/her investment according to the recommended portfolio on a quarterly basis with no friction cost.

## V. RESULT AND DISCUSSION

We construct "Buy" and "Sell" portfolios for each quarter in the test set based on the predictions provided by FNN and ANFIS models. The mean and standard deviation of the real quarterly relative return of these equal-weight portfolios are then computed. The real quarterly relative returns of an equal-weight portfolio which includes all 70 stocks are also computed as a benchmark for comparison. The relative return which compounds the 18 quarters in the test set was calculated in the end. Our observations based on the experimental results are as follows:

- Both "Buy" portfolios constructed using FNN and ANFIS models outperform the full sample universe of 70 stocks in terms of mean quarterly relative return by a significant margin. On the other hand, both "Sell" portfolios underperform the benchmark. The compound results confirm this finding. We can land a safe conclusion that both FNN and ANFIS are able to predict, with some degree of accuracy, the near-term winners and losers from a universe of stocks based on the stocks' most recent fundamental financial ratios. The results obtained challenge both the weak and the semi-strong form of EMH.

- FNN outperforms ANFIS in constructing both "Buy" and "Sell" portfolios in terms of mean quarterly relative return. In terms of compound relative return, FNN also dominates ANFIS by a large margin.

- The standard deviations of quarterly relative returns for the selected portfolios are higher than that of the benchmark. This means the selected portfolios are more volatile than the benchmark. Excessive volatility would diminish compound return in the long term. However, the simulated compound returns for the selected portfolios are significantly higher than that of the benchmark, especially for FNN. This confirms the conclusion that both FNN and ANFIS have the ability to separate winners from losers based on fundamental ratios.

- Both FNN and ANFIS are better at identifying losers than identifying winners by a small margin. More investigation is required to find the reasons behind such a phenomenon.

The complete results for both models are illustrated in Table I and II. Fig. 3 shows the compound relative return of the portfolios constructed by both models for the test period.

TABLE I
CONSTRUCTED 'BUY' PORTFOLIO RELATIVE RETURN

|  | Mean | STD | Compound |
|---|---|---|---|
| FNN | 0.72% | 3.92 | 12.34% |
| ANFIS | 0.27% | 4.07 | 3.45% |
| Full Sample | -0.02% | 3.55 | -1.35% |

TABLE II
CONSTRUCTED 'SELL' PORTFOLIO RELATIVE RETURN

|  | Mean | STD | Compound |
|---|---|---|---|
| FNN | -0.83% | 3.87 | -14.97% |
| ANFIS | -0.34% | 4.03 | -7.19% |
| Full Sample | -0.02% | 3.55 | -1.35% |

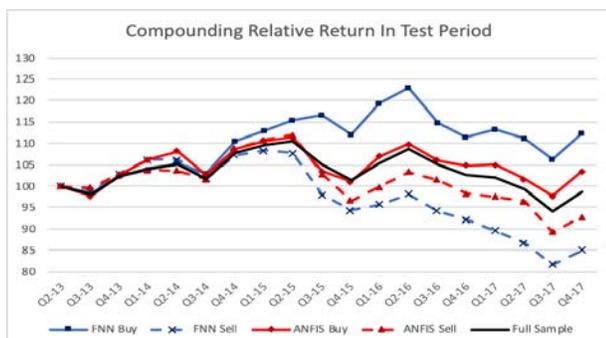

Fig. 3. Compound Relative Return for Portfolios Selected by NNs for Test Period

## VI. CONCLUSION AND FUTURE WORK

In this research, we explored applying FNN and ANFIS for stock return prediction based on a stock's fundamental financial ratios rather than commonly used technical indicators. An individual stock's relative return with respect to the DJIA index was used for model training and testing rather than absolute return. Predicted relative returns of 70 stocks were then ranked, and "Buy" and "Sell" portfolios, which consist of the top 30 and bottom 30 stocks respectively, were selected for each quarter. The result of our experiment reveals that "Buy" portfolios and "Sell" portfolios selected by FNN and ANFIS are able to respectively outperform and underperform the full sample universe. Moreover, FNN noticeably outperforms ANFIS in portfolio selection.

As for future work, iterative feature selection can be employed to reduce the number of features and potentially improve prediction accuracy. The sample universe may be expanded and the time span of data may be manipulated to further validate the findings of this research.

REFERENCES

[1] E. F. Fama, The behavior of stock market prices, *The Journal of Business, 38*, 34-105, 1965.
[2] B.M. Blau, The volatility of exchange rates and the non-normality of stock returns, *Journal of Economics and Business*, 91, 41-52, 2017.
[3] E. Guresen, G. Kayakutlu, TU Daim, Using artificial neural network models in stock market index prediction, *Expert Systems with Applications*, 2011.
[4] G.S. Atsalakis, K.P. Valavanis, Forecasting stock market short-term trends using a neuro-fuzzy based methodology, *Expert Systems with Applications, 36,* 10696-10707, 2009.
[5] J Patel, S. Shah, P. Thakkar, K. Kotecha, Predicting stock and stock price index movement using trend deterministic data preparation and machine learning techniques, *Expert Systems with Applications, 42(1), 259-268*, 2015.
[6] S. Bekiros, D. Georgoutsos, Evaluating direction-of-change forecasting: Neurofuzzy models vs. neural networks, *Mathematical and Computer Modeling, 46, 38-46*, 2007.
[7] M.K. Farahani, S. Mehralian, Comparison between artificial neural network and neuro-fuzzy for gold price prediction, *13th Iranian Conf. on Fuzzy Systems,* 2013.
[8] K.Y. Shen, G.H. Tzeng, Combined soft computing model for value stock selection based on fundamental analysis, *Applied Soft Computing, 37,* 142-155, 2015.
[9] S.G. Eakins, S.R. Stansell, Can value-based stock selection criteria yield superior risk-adjusted returns: an application of neural networks, *International Review of Financial Analysis,* 12, 83-97, 2003.
[10] T.S. Quah, B. Srinivasan, Improving returns on stock investment through neural network selection, *Expert Systems with Applications,* 17, 295-301, 1999
[11] T.S. Quah, DJIA stock selection assisted by neural network, *Expert Systems with Applications*, 35, 50-58, 2008.
[12] F. Black, The trouble with econometrics models. *Financial Analysis Journal*, 4(5), 75-87, 1982.
[13] Y. Hu, K. Liu, X. Zhang, L. Su, E.W.T. Ngai, M. Liu, Application of evolutionary computation for rule discovery in stock algorithmic trading: A literature review. *Applied Soft Computing, 36,* 534-551, 2015.
[14] G.S. Atsalakis, K.P. Valavanis, Surveying stock market forecasting techniques-part II:soft computing methods. *Expert System with Applications, 36(3)* 5932-5941, 2009.
[15] S. Haykin, Neural Networks: A Comprehensive Foundation, 2rd edition, Prentice Hall, pp. 183-195, 1998
[16] J.R. Jang, ANFIS: adaptive-network-based fuzzy inference system, in *IEEE Trans. on Systems, Man, and Cybernetics*, 23(3), 665-685, 1993.
[17] S&P 100 Fact Sheet, https://us.spindices.com/indices/equity/sp-100, Nov. 30, 2018 [Dec. 10, 2018].
[18] Stockpup, http://www.stockpup.com [July. 1, 2018].
[19] J.W. Graham, Missing data analysis: making it work in the real world, *Annual Review of Psychology*, 60, 549-576, 2009.